\documentclass[10pt]{iopart}
\usepackage{cite}
\usepackage{graphicx}

\newlength{\figurewidth}
\begin{document}

\paper{Neural cryptography with queries}
\author{Andreas Ruttor$^1$, Wolfgang Kinzel$^1$ and Ido Kanter$^2$}
\address{$^1$ Institut f\"ur Theoretische Physik und Astrophysik,
  Universit\"at W\"urzburg, Am Hubland, 97074 W\"urzburg, Germany}
\address{$^2$ Department of Physics, Bar Ilan University, Ramat Gan
  52900, Israel}
\eads{\mailto{andreas.ruttor@physik.uni-wuerzburg.de},
  \mailto{wolfgang.kinzel@physik.uni-wuerzburg.de} and
  \mailto{kanter@mail.biu.ac.il}}

\begin{abstract}
  Neural cryptography is based on synchronization of tree parity
  machines by mutual learning. We extend previous key-exchange
  protocols by replacing random inputs with queries depending on the
  current state of the neural networks. The probability of a
  successful attack is calculated for different model parameters using
  numerical simulations. The results show that queries restore the
  security against cooperating attackers. The success probability can
  be reduced without increasing the average synchronization time.
\end{abstract}

\begin{indented}
\raggedright
\item \textbf{Keywords:} stochastic processes (theory), neuronal
  networks (theory), new applications of statistical mechanics
\end{indented}

\section{Introduction}

Neural cryptography \cite{Kanter:2002:SEI, Kinzel:2003:DGI} is a
method for generating secret information over a public channel. Before
two partners A and B can exchange a secret message over a public
communication channel they have to agree on a secret encryption key.
Using number theoretic methods, such keys can be constructed over
public channels without previous secret agreements of the two partners
\cite{Stinson:1995:CTP}. Any details of the algorithm as well as the
complete information passed between the partners are known to a
possible attacker E; nevertheless the final key is secret, it is known
to the two partners A and B only, and an opponent E with limited
computer power cannot calculate the key. The method is based on the
computational difficulty of factorizing large numbers or calculating
the discrete logarithm of large numbers \cite{Stinson:1995:CTP}.

Recently it has been demonstrated that secret keys can be generated by
a completely different method \cite{Kanter:2002:SEI}, as well. This
method is based on synchronization of neural networks by mutual
learning \cite{Metzler:2000:INN, Kinzel:2000:DIN}. The secret key is
generated by the dynamics of a complex physical process, namely the
competition between stochastic attractive and repulsive forces which
act on the weights of the two neural networks of the partners A and B.
Two dynamical systems which synchronize by mutual signals have an
advantage over an attacker E which can only synchronize by listening
to the exchanged signals \cite{Kinzel:2003:DGI}. Finally the key is
taken as the synchronized weights of the two networks A and B.

The security of neural cryptography is still being debated
\cite{Klimov:2003:ANC, Kinzel:2002:INN, Kinzel:2002:NC,
  Mislovaty:2002:SKE, Kanter:2002:TNN, Shacham:2003:CAN}. Since the
method is based on a stochastic process, there is a small chance that
an attacker synchronizes to the key, as well. However, it has been
found that the model parameter $L$ (the synaptic depth defined below)
determines the security of the system: the success probability of the
attacker decreases exponentially with $L$ while the synchronization
time, i.e.~the amount of effort for agreeing on a key, increases by
$L^2$, only. Hence, by increasing the value of $L$ the security of
neural cryptography can be increased to any desired level
\cite{Mislovaty:2002:SKE}.

These scaling laws are not obvious at all. Neural cryptography is
based on a subtle difference between bi-directional and
uni-directional couplings of a stochastic process. Our understanding
of these processes is still incomplete. Hence it is still possible
that a clever algorithm may destroy the security of the method.

In fact, there is a special algorithm of neural cryptography, the
Hebbian rule defined below, which allows a clear determination of
scaling laws with respect to the parameter $L$. Recently it has been
shown that a special attack based on the majority of an ensemble of
attacking networks destroys the security of the method: the success
probability is constant for large values of $L$
\cite{Shacham:2003:CAN}. These results are the motivation for the
present investigation. We develop a new kind of learning rule which
restores the security: the success probability decreases exponentially
with $L$. This new rule uses \emph{learning by queries}
\cite{Kinzel:1990:ING} which is a well-known principle in the theory
of learning by examples \cite{Engel:2001:SML}. It is based on
exchanging inputs between A and B which are correlated to the weight
vectors of the two networks.

It should be mentioned that there are a few other rules (anti-Hebbian
or random walk) with lower success probabilities for which the scaling
properties with respect to the parameter $L$ cannot be found, yet.
This means that we still do not know whether the majority attack
destroys the security of those algorithms, as well. Nevertheless it is
important to develop a rule which restores security for the Hebbian
rule where the majority attack is clearly successful.

Our findings allow us to reach the conclusion that for all algorithms
suggested so far the scaling laws for the success probability hold:
the security of neural cryptography can be increased to any desired
level.

\section{Neural cryptography}
\label{sec:nc}

\begin{figure}
  \centering
  \includegraphics[width=\figurewidth]{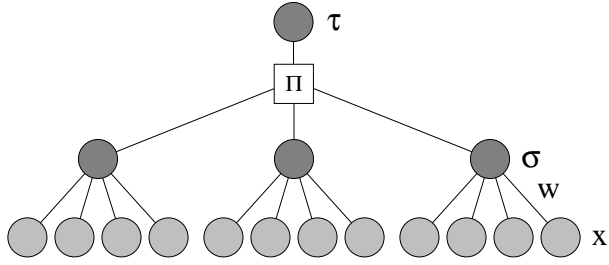}
  \caption{A tree parity machine with $K=3$ and $N=4$.}
  \label{fig:tpm}
\end{figure}

In this section we repeat the definition of neural cryptography. Each
of the two partners A and B uses a special neural network called tree
parity machine (TPM). As shown in \fref{fig:tpm}, a TPM consists of
$K$ hidden units $\sigma_k$ with weight vectors $\bi{w}_k$ and input
vectors $\bi{x}_k$. The components of the input vectors are binary and
the weights are discrete numbers with depths $L$,
\begin{equation}
  x_{k,j} \in \{-1,+1\}, \quad w_{k,j} \in \{-L,-L+1,...,L-1,L\},
\end{equation}
where the index $j = 1, \dots, N$ denotes the elements of each vector
and $k = 1, \dots, K$ the hidden units. The outputs of these neurons
are defined by the scalar product of inputs and weights,
\begin{equation}
  \sigma_k = \mathrm{sgn} (\bi{w}_k \cdot \bi{x}_k) \, .
\end{equation}

The final output bit of each TPM is defined by the product of the
hidden units,
\begin{equation}
  \tau= \prod_{k=1}^{K} \sigma_k \, .
\end{equation}

Both partners A and B initialize their weight vectors by means of
random numbers before the training period starts. At each time step
$t$ a public input vector is generated and the bits $\tau^\mathrm{A}$
and $\tau^\mathrm{B}$ are exchanged over the public channel. In the
case of identical output bits, $\tau^\mathrm{A}=\tau^\mathrm{B}$, each
TPM adjusts those of its weights for which the hidden unit is
identical to the output, $\sigma_k^\mathrm{A/B} = \tau^\mathrm{A/B}$.
These weights are adjusted according to a given learning rule. Here we
consider the Hebbian rule
\begin{equation}
  \label{eq:hebb}
  \bi{w}^\mathrm{A/B}_k(t+1) = \bi{w}^\mathrm{A/B}_k(t) +
  \tau^\mathrm{A/B} \bi{x}_k \, .
\end{equation}
After some time $t_\mathrm{sync}$ the two partners are synchronized,
$\bi{w}_k^\mathrm{A}(t)=\bi{w}_k^\mathrm{B}(t)$, and the communication
is stopped. Then the common weight vector is used as a key to encrypt
secret messages.

Note that any possible attacker E knows as much about the process as A
knows about B and vice versa. But E has some disadvantage with respect
to A and B: it can only listen to the communication and cannot
influence the dynamics of the weights in A's and B's neural networks
\cite{Kinzel:2002:TIN, Kinzel:2003:DGI}. It turns out that this
difference determines the security of the crypto-system.

It seems obvious that a successful attacker should also use a TPM with
similar training rules. In fact, it is advantageous to use many
networks for an attack. Since the method is stochastic (due to the
random input) the attacker may synchronize as well, but with a low
probability $P_\mathrm{E}$. The security of neural cryptography is
related to the fact that $P_\mathrm{E}$ decreases exponentially with
the value of $L$ \cite{Mislovaty:2002:SKE}.

In all previous learning rules the sequence of input vectors
$\bi{x}_k(t)$ was generated by a public random number generator. Here
we propose to take queries, i.e.~input vectors which are correlated
with the present weight vector $\bi{w}_k(t)$. At odd (even) time steps
the partner A (B) is generating an input vector which has a certain
overlap to its weights $\bi{w}_k^\mathrm{A}$ ($\bi{w}_k^\mathrm{B}$).
It turns out that queries improve the security of the system.

\section{Queries}

The process of synchronization itself can be described by standard
order parameters which are also used for the analysis of on-line
learning \cite{Engel:2001:SML}. These order parameters are
\begin{eqnarray}
  Q_k^m     &=& \frac{1}{N} \bi{w}_k^m \cdot \bi{w}_k^m \, , \\
  R_k^{m,n} &=& \frac{1}{N} \bi{w}_k^m \cdot \bi{w}_k^n \, ,
\end{eqnarray}
where the indices $m,n \in \{A,B,E\}$ denote A's, B's, or E's TPM.
The distance between two corresponding hidden units is defined by the
(normalized) overlap
\begin{equation}
  \rho_k^{m,n} = \frac{\bi{w}_k^m \cdot \bi{w}_k^n}{\sqrt{\bi{w}_k^m
      \cdot \bi{w}_k^m} \sqrt{\bi{w}_k^n \cdot \bi{w}_k^n}} =
  \frac{R_k^{m,n}}{\sqrt{Q_k^m \, Q_k^n}} \, .
\end{equation}
The maximum value $\rho_k=1$ is reached for fully synchronized hidden
units (zero distance), while uncorrelated weight vectors have zero
overlap (maximal distance).

The distance between two corresponding hidden units decreases, if the
learning rule \eref{eq:hebb} is applied to both of them using the same
input vector $\bi{x}_k$. Thus coordinated moves of the weights have an
attractive effect in the process of synchronization
\cite{Klimov:2003:ANC, Ruttor:2004:SRW}.

But changing the weights in only one hidden unit increases the average
distance \cite{Klimov:2003:ANC}. These repulsive steps between two
corresponding hidden units can only occur if their output bits are
different \cite{Ruttor:2004:NCF}. The probability for this event is
given by the well-known generalization error of the perceptron
\cite{Engel:2001:SML}
\begin{equation}
  \label{eq:epsilon}
  \epsilon_k = \frac{1}{\pi} \arccos \rho_k \, ,
\end{equation}
which depends on the overlap $\rho_k$ between the weight vectors of
corresponding hidden units. For an attacker using the same learning
rule as A and B, repulsive steps in the $k$th hidden unit occur with
probability $p_\mathrm{r} = \epsilon_k$, as E cannot influence the
process of synchronization.

In contrast, the partners update the weights in their TPMs only if
$\tau^\mathrm{A} = \tau^\mathrm{B}$. In the case of identical
generalization error, $\epsilon_k=\epsilon$, we find for $K=3$ that
repulsive steps in a hidden unit of B's neural network occur with the
probability \cite{Ruttor:2004:NCF}
\begin{equation}
  p_\mathrm{r} = \frac{2 (1-\epsilon) \epsilon^2}{(1-\epsilon)^3 + 3
    (1-\epsilon) \epsilon^2} \leq \epsilon \, .
\end{equation}
So $p_\mathrm{r}$ is lower for synchronization than for learning and
the partners partially avoid repulsive steps. This advantage makes
neural cryptography feasible and prevents successful attacks, which
are only based on simple learning.

But E can assign a confidence level to each output
$\sigma^\mathrm{E}_k$ of its hidden units. For this task the local
field
\begin{equation}
  h_k = \frac{1}{\sqrt{N}} \bi{w}_k \cdot \bi{x}_k \, ,
\end{equation}
is used as additional information. Then the prediction error, the
probability of different output bits for an input vector $\bi{x}_k$
inducing a local field $h_k$, is given by \cite{Ein-Dor:1999:CPN}
\begin{equation}
  \label{eq:error}
  \epsilon(\rho_k,h_k) = \frac{1}{2} \left[ 1 - \mathrm{erf} \left(
      \frac{\rho_k}{\sqrt{2 (1 - \rho_k^2)}} \frac{|h_k|}{\sqrt{Q_k}}
    \right) \right] \, .
\end{equation}
The function $\epsilon(\rho_k,h_k)$ reaches a maximum of
$\epsilon(\rho_k,0)=0.5$ for $h_k=0$. In this case $\bi{x}_k$ is
perpendicular to $\bi{w}_k$ and the neural network has no information
about this example. But for increasing $|h_k|$ the confidence of
prediction rises and $\epsilon(\rho_k,h_k)$ shrinks.

This effect is essential for the \emph{Geometric Attack}
\cite{Klimov:2003:ANC}, which is---up to now---the most successful
method for a single attacking neural network with a structure
identical to A's and B's \cite{Mislovaty:2002:SKE}. Here E trains its
own TPM, with the examples, input vectors and output bits, transmitted
by the two partners. If $\tau^\mathrm{E} = \tau^\mathrm{A}$, the
attacker applies the same learning rule as A and B. But for
$\tau^\mathrm{E} \not= \tau^\mathrm{A}$ E knows that at least one of
the hidden units has made a wrong prediction and searches for the unit
$k$ with the minimum absolute value $|h_k|$ of the local field.
Because this hidden unit has the lowest confidence of prediction, its
output $\sigma^\mathrm{E}_k$ is most likely to be different from
$\sigma^\mathrm{A}_k$. Therefore the attacker inverts both
$\sigma^\mathrm{E}_k$ and $\tau^\mathrm{E}$. Afterwards the usual
learning rule can be applied. As this geometric attack method reduces
the frequency of repulsive steps, E increases its success probability
$P_\mathrm{E}$ by taking the local field into account.

But the partners can influence the local field. Instead of using
random inputs, they are able to select input vectors with a fixed
$|h_k|$ (queries \cite{Kinzel:1990:ING}) in their own hidden units.
This essentially modifies the functional dependency between overlap
$\rho_k$ and frequency of repulsive steps $p_\mathrm{r}$. In this case
\eref{eq:error} determines the probability of different output bits
instead of \eref{eq:epsilon}.

Note that the chosen absolute local field $|h_k|$ for synchronization
with queries is lower than the average value
\begin{equation}
  \langle |h_k| \rangle = \sqrt{2 Q_k / \pi} \approx 0.8 \sqrt{Q_k}
\end{equation}
observed for random inputs. Therefore the overlap
\begin{equation}
  \rho_{k,\mathrm{in}} = \frac{\bi{w}_k \cdot \bi{x}_k}{\sqrt{\bi{w}_k
      \cdot \bi{w}_k} \sqrt{\bi{x}_k \cdot \bi{x}_k}} =
  \frac{1}{\sqrt{N}} \frac{h_k}{\sqrt{Q_k}}
\end{equation}
between input vector and weight vector converges to zero in the limit
$N \rightarrow \infty$, even if queries with $0 < |h_k| < \infty$ are
used.

\begin{figure}
  \centering
  \includegraphics[width=\figurewidth]{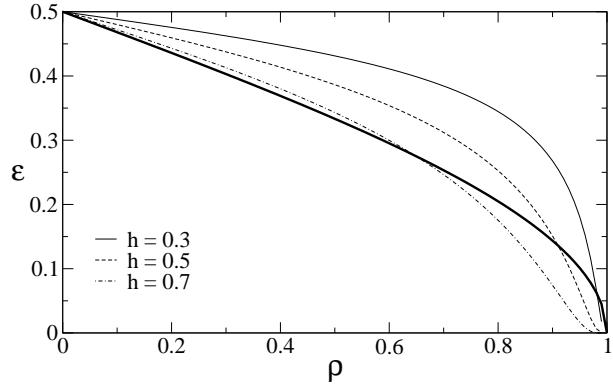}
  \caption{Prediction error $\epsilon(\rho,h)$ as a function of the
    overlap $\rho$ for different values of the local field. The
    generalization error $\epsilon$ is shown as thick line. We assumed
    $Q=1$ for this diagram.}
  \label{fig:epsilon}
\end{figure}

\Fref{fig:epsilon} shows the prediction error $\epsilon(\rho,h)$ for
queries, which induce the local field $h$. Compared to the case of
random inputs, $\epsilon(\rho,h)$ is increased for small overlap and
decreased for nearly synchronized neural networks. By selecting
queries with different local fields, A and B can regulate this effect.

As learning is slower than synchronization, $\rho^\mathrm{AE}$ is
typically smaller than $\rho^\mathrm{AB}$. In this situation queries
with small values of the parameter $h$ increase the frequency of
repulsive steps for the attacker without affecting the process of
synchronization too much.

\section{Synchronization}

In order to integrate queries into the neural key-exchange protocol
\cite{Kanter:2002:SEI}, only the generation of the inputs has to be
changed. Instead of choosing completely random $x_{k,j}$, both
partners alternately ask each other questions. In each time step
either A or B uses the algorithm described in \ref{sec:qgen} to
generate $K$ input vectors $\bi{x}_k$, which result in
$h_k^\mathrm{A/B} \approx \pm H$. Then these queries are sent to the
other partner and both calculate the output of their TPMs. After the
exchange of $\tau^\mathrm{A}$ and $\tau^\mathrm{B}$, the Hebbian
learning rule is used to update the weights as described in
\sref{sec:nc}. This leads to synchronization after $t_\mathrm{sync}$
steps.

\begin{figure}
  \centering
  \includegraphics[width=\figurewidth]{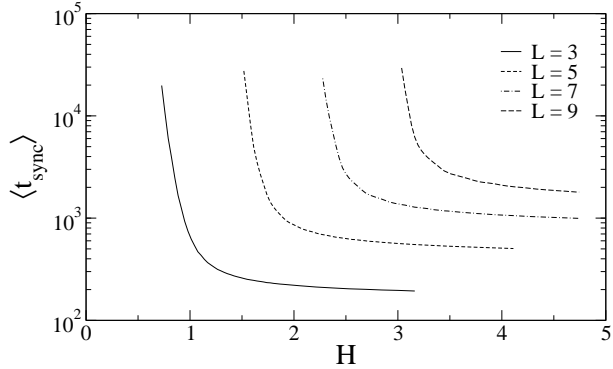}
  \caption{Synchronization time of two TPMs with $K=3$
    and $N=1000$, averaged over $10\,000$ simulations.}
  \label{fig:sync1}
\end{figure}

As shown in \fref{fig:sync1}, $\langle t_\mathrm{sync} \rangle$
diverges for $H \rightarrow 0$. In this limit the prediction error
$\epsilon(\rho_k,H)$ reaches $0.5$ independent of $\rho_k$. Therefore
the effect of the repulsive steps inhibits synchronization. But the
choice of $H$ does not influence $\langle t_\mathrm{sync} \rangle$
much, as long as this quantity is large enough.

The dependence on $L$ of $\langle t_\mathrm{sync} \rangle$ is caused
by two effects:
\begin{itemize}
\item The dynamics of each weight can be described as random walk with
  reflecting boundaries if the control signals $\sigma_k$ and $\tau$
  are neglected. Calculations for this simplified model of neural
  synchronization show that $\langle t_\mathrm{sync} \rangle$ scales
  proportional to $L^2$ \cite{Ruttor:2004:SRW}. This behavior has been
  observed in neural cryptography with random inputs
  \cite{Mislovaty:2002:SKE}, too.
\item If queries are used, the probability of repulsive steps
  $p_\mathrm{r}$ depends not only on the overlap $\rho_k$, but also on
  the quantity $|h_k| / \sqrt{Q_k}$. Assuming uniformly distributed
  weights, we find
  \begin{equation}
    \langle Q_k \rangle = \left\langle \frac{1}{N} \bi{w}_k \cdot
      \bi{w}_k \right\rangle = \frac{1}{3} L (L + 1) \sim \frac{1}{3}
    L^2
  \end{equation}
  for the expectation value of the order parameter $Q_k$. Hence we
  have to increase $H$ proportional to the synaptic depth $L$, if we
  want to observe the same frequency of repulsive steps.
\end{itemize}
Using both $\langle t_\mathrm{sync} \rangle \propto L^2$ and $H
\propto L$ we can rescale $\langle t_\mathrm{sync} \rangle_{H,L}$ in
order to obtain functions $f_L(\alpha)$, which are nearly independent
of the synaptic depth in the case $L \gg 1$:
\begin{equation}
  \langle t_\mathrm{sync} \rangle = L^2 f_L \! \left( \frac{H}{L}
  \right) \, .
\end{equation}

\begin{figure}
  \centering
  \includegraphics[width=\figurewidth]{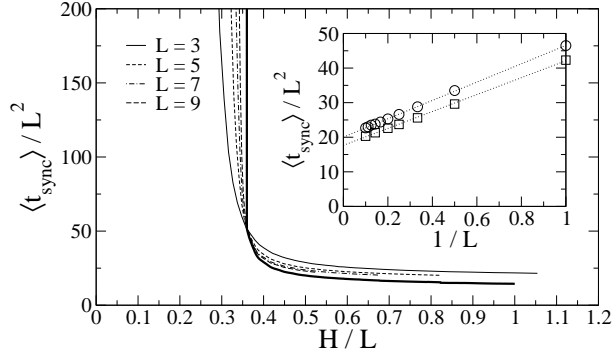}
  \caption{Scaling behavior of the synchronization time. The thick
    curve denotes the universal function $f(\alpha)$ defined in
    \eref{eq:f}. It has been obtained by finite size scaling, which is
    shown in the inset for $\alpha = 0.5$ ($\circ$) and $\alpha = 0.6$
    (\opensquare).}
  \label{fig:sync2}
\end{figure}

In \fref{fig:sync2} we have plotted these functions for different
values of $L$. It is clearly visible that $f_L(\alpha)$ converges to a
universal scaling function $f(\alpha)$ in the limit $L \rightarrow
\infty$:
\begin{equation}
  \label{eq:f}
  f(\alpha) = \lim_{L \rightarrow \infty} f_L(\alpha) \, .
\end{equation}
Additionally, we find that the distance $|f_L(\alpha) - f(\alpha)|$
shrinks proportional to $L^{-1}$. Therefore we can use finite size
scaling to determine $f(\alpha)$, which is shown in \fref{fig:sync2},
too.

\begin{figure}
  \centering
  \includegraphics[width=\figurewidth]{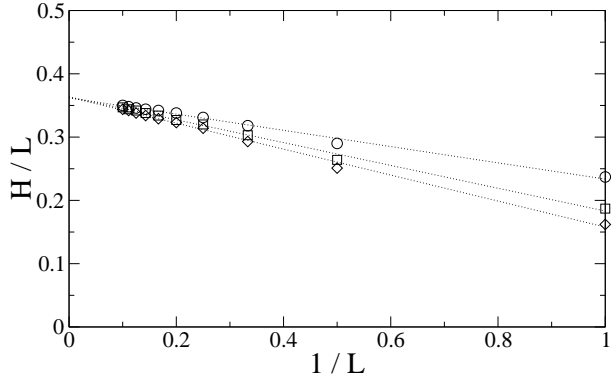}
  \caption{Extrapolation of $f_L^{-1}$ to $L \rightarrow
    \infty$. Symbols denote the values extracted from \fref{fig:sync2}
    for the average synchronization times $100 L^2$ ($\circ$), $150
    L^2$ (\opensquare) and $200 L^2$ (\opendiamond).}
  \label{fig:sync3}
\end{figure}

This function diverges for $\alpha < \alpha_\mathrm{c}$. We estimate
$\alpha_\mathrm{c} \approx 0.36$ for $K=3$ and $N=1000$ by
extrapolating the inverse function $f_L^{-1}$ as shown in
\fref{fig:sync3}. Consequently, synchronization is only achievable for
$H > \alpha_\mathrm{c} L$ in the limit $L \rightarrow \infty$.

\section{Security}

Up to now, the most successful attack on neural cryptography is the
\emph{Majority Flipping Attack} \cite{Shacham:2003:CAN}, which is an
extension of the \emph{Geometric Attack}. Instead of a single neural
network, E uses an ensemble of $M$ TPMs. At the beginning, the weight
vectors of all attacking networks are chosen randomly, so that the
average initial overlap between them is zero. Like A and B, the
attacker only updates the weights in time steps with $\tau^\mathrm{A}
= \tau^\mathrm{B}$. If the output $\tau^{\mathrm{E},m}$ of the $m$th
attacking network disagrees with $\tau^\mathrm{A}$, the hidden unit
with the smallest absolute value $|h^{\mathrm{E},m}_k|$ of the local
field is selected. Then the output bits $\sigma^{\mathrm{E},m}_k$ and
$\tau^{\mathrm{E},m}$ are inverted. Afterwards E counts the internal
representations $(\sigma^{\mathrm{E},m}_1, \dots,
\sigma^{\mathrm{E},m}_K)$ and selects the most common one. This
majority vote is then adopted by all attacking networks for the
application of the learning rule.

Because of these identical updates, E's neural networks become
correlated \cite{Shacham:2003:CAN}, which reduces the efficiency of
the \emph{Majority Flipping Attack}. In order to slow down this
effect, we use the modifications proposed in \cite{Shacham:2003:CAN}:
\begin{itemize}
\item The majority vote is only considered in even time steps.
  Otherwise E updates the weights according to the internal
  representation of the particular neural network.
\item At the beginning of the synchronization, only the
  \emph{Geometric Attack} is applied by the attacker. But instead of
  waiting until $t > \frac{1}{3} \, t_\mathrm{sync}$ as suggested in
  \cite{Shacham:2003:CAN}, E starts with the \emph{Majority Flipping
    Attack} after 100 steps.
\end{itemize}

For the neural key-exchange protocol with random inputs and Hebbian
learning rule, the success probability $P_\mathrm{E}$ of this method
reaches a constant non-vanishing value in the limit $L \rightarrow
\infty$ \cite{Shacham:2003:CAN}. Therefore it breaks the security of
this type of neural cryptography.

\begin{figure}
  \centering
  \includegraphics[width=\figurewidth]{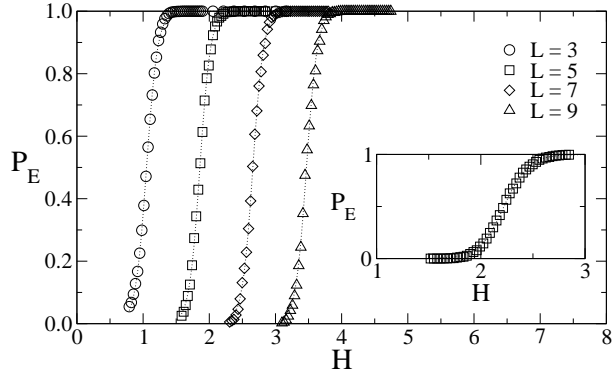}
  \caption{Success probability of the \emph{Majority Flipping Attack}
    as a function of $H$. Symbols denote the results obtained in
    $10\,000$ simulations for $K=3$, $M=100$ and $N=1000$. The inset
    shows the success of a \emph{Geometric Attack} for $K=3$, $L=5$
    and $N=1000$.}
  \label{fig:success}
\end{figure}

In contrast, if the two partners A and B use queries for the neural
key exchange, the success probability strongly depends on the
parameter $H$. This can be used to regain security against the
\emph{Majority Flipping Attack}. As shown in \fref{fig:success} a
Fermi-Dirac distribution
\begin{equation}
  \label{eq:fermi}
  P_\mathrm{E} = \frac{1}{1 + \exp(-\beta (H - \mu))}
\end{equation}
with two parameters $\beta$ and $\mu$ is a suitable fitting function
for describing $P_\mathrm{E}$ as a function of $H$. \Eref{eq:fermi} is
also valid for the \emph{Geometric Attack}, which is a special case of
the \emph{Majority Flipping Attack} ($M=1$).

\begin{figure}
  \centering
  \includegraphics[width=\figurewidth]{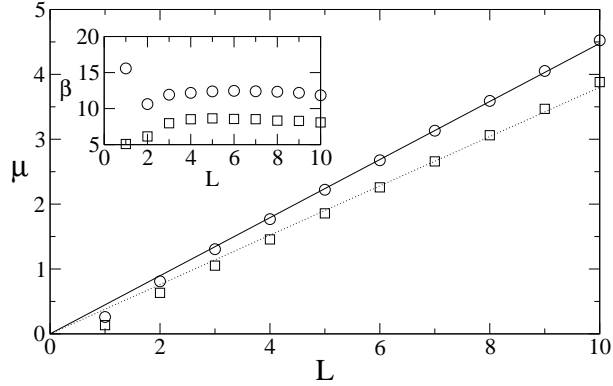}
  \caption{Parameters $\mu$ and $\beta$ as a function of the synaptic
    depth $L$. Symbols denote the results of fits using
    \eref{eq:fermi} for the \emph{Geometric Attack} ($\circ$) and the
    \emph{Majority Flipping Attack} with $M=100$ (\opensquare). From
    these values we obtain $\alpha_\mathrm{s,flip} \approx 0.45$
    (\full) and $\alpha_{\mathrm{s},M=100} \approx 0.38$ (\dotted)
    according to \eref{eq:prop}.}
  \label{fig:mfga}
\end{figure}

\Fref{fig:mfga} shows the dependence on $L$ of the fit parameter $\mu$
for both attacks. Aside from finite size effects for small values of
$L$, this parameter is proportional to the synaptic depth of the TPMs:
\begin{equation}
  \label{eq:prop}
  \mu = \alpha_\mathrm{s} \, L \, .
\end{equation}
Obviously, the quantity $\alpha = H / L$ not only determines the
synchronization time but also the success of an attack. These effects
are caused by the modification of $p_\mathrm{r}$ due to the use of
queries. The other parameter $\beta$ is nearly constant for $L > 3$.
So both $\alpha_\mathrm{s}$ and $\beta$ are independent of the chosen
parameters $H$ and $L$, but depends on the attacker's method.

From these results we can also deduce the scaling behavior of
$P_\mathrm{E}$ as a function of the synaptic depth. For both the
\emph{Majority Flipping Attack} and the \emph{Geometric Attack} we
obtain
\begin{equation}
  \label{eq:result}
  P_\mathrm{E} = \frac{1}{1 + \exp(\beta (\alpha_\mathrm{s} - \alpha)
    L)}
\end{equation}
for synchronization with queries and $H = \alpha L$. In the limit $L
\rightarrow \infty$, the asymptotic behavior of $P_\mathrm{E}$ is
given by
\begin{equation}
  P_\mathrm{E} \sim \mathrm{e}^{-\beta (\alpha_\mathrm{s} - \alpha) L}
\end{equation}
as long as $\alpha < \alpha_\mathrm{s}$. Here the success probability
$P_\mathrm{E}$ decreases exponentially with increasing synaptic depth,
\begin{equation}
  \label{eq:scale}
  P_\mathrm{E} \propto \mathrm{e}^{-y L} \, ,
\end{equation}
which is also observed in the case of random inputs, if E uses the
\emph{Geometric Attack} \cite{Mislovaty:2002:SKE}.

Consequently, the neural key exchange with queries is secure against
both attacks. An attacker using the \emph{Majority Flipping Attack}
only decreases the value of $y = \beta (\alpha_\mathrm{s} - \alpha)$,
because $\alpha_\mathrm{s,majority} < \alpha_\mathrm{s,flip}$, but
does not change the exponential scaling law \eref{eq:scale}.

\begin{figure}
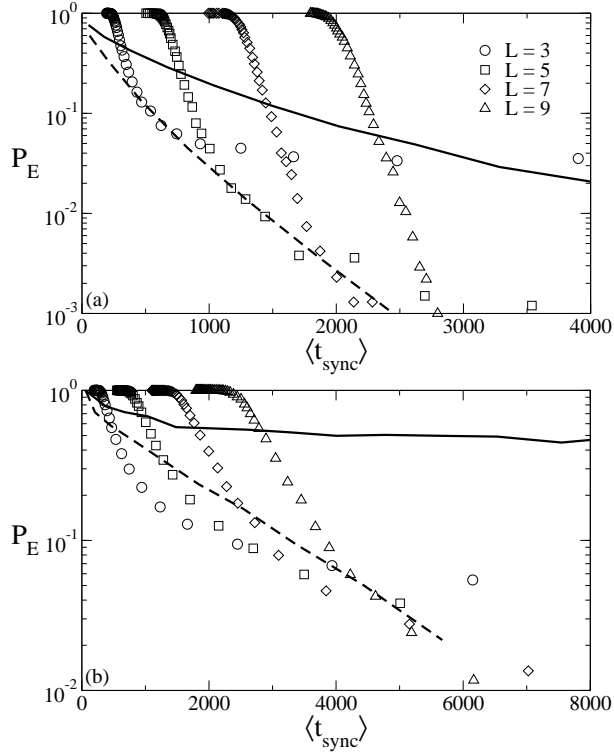

  \centering
  \includegraphics[width=\figurewidth]{gresult.eps}
  \includegraphics[width=\figurewidth]{mresult.eps}
  \caption{Success probability as a function of the average
    synchronization time for $K=3$, $N=1000$ and different values of
    $L$ and $H$. Part (a) shows the result for the \emph{Geometric
      Attack} and part (b) for the \emph{Majority Flipping Attack}
    with $M=100$ attacking neural networks. The solid curve in each
    graph represents the success probability for neural cryptography
    with random inputs and the dashed line marks $H = \alpha_c \, L$.}
  \label{fig:result}
\end{figure}

For practical aspects of security, however, one has to look at the
function $P_\mathrm{E}(\langle t_\mathrm{sync} \rangle)$, which is
shown in \fref{fig:result}. Here we find that queries enhance the
security of the neural key-exchange protocol a lot for given
synchronization time. There is an optimal value of $H$ associated with
each $L$, which lies on the envelope of all functions
$P_\mathrm{E}(\langle t_\mathrm{sync} \rangle)$.

\Fref{fig:result} also shows that A and B can achieve higher security
against attacks than predicted by \eref{eq:result} with $\alpha =
\alpha_\mathrm{c}$ as long as $L$ is not to large. This phenomenon is
caused by finite size effects which enable synchronization even for $H
< \alpha_\mathrm{c} \, L$. But, of course, this does not work for $L
\gg 1$. Therefore the envelope and $P_\mathrm{E}(\langle
t_\mathrm{sync} \rangle)$ for $H = \alpha_\mathrm{c} \, L$ converge
asymptotically in the limit $L \rightarrow \infty$.

Queries reveal additional information about the weight vectors of A's
and B's neural networks. However, an attacker E cannot benefit from
this information, since for a given value of $H$, there is still an
exponential large number of weight vectors $\bi{w}_k$, which are
consistent with a given query. As an example, there are $2.8 \times
10^{129}$ possible weight vectors for $L=10$, $N=100$, and $h_k=10$.
Because of this large number, E cannot gain useful information from
queries.

\section{Conclusions}

Neural cryptography is a subtle competition between interaction and
learning of neural networks. Two neural networks A and B exchange some
information about their states over a public channel. The amount of
information has to be so high that the two networks A and B can
synchronize. But it has to be so low that an attacker E can only
synchronize with a low probability which can be decreased to an
arbritrary low value.

We have shown that increasing the amount of exchanged information can
be of advantage for cryptography. We have included queries in the
training process of the neural networks. This means that alternately A
and B are generating an input which is correlated with its state and A
or B is asking the partner for the corresponding output bit. The
overlap between input and weight vector is so low that the additional
information does not reveal much about the internal states. But
queries introduce a mutual influence between A and B which is not
available to an attacking network E. In addition the method obtains a
new (public) parameter which can be adjusted to give optimal security.

We have applied this new method to the case of the Hebbian training
rule which was successfully attacked using the majority of an ensemble
of attackers. We find that queries restore the security of the method:
the probability of a successful naive majority attack can be decreased
to any desired level.

\appendix

\section{Generation of queries}
\label{sec:qgen}

In this appendix we describe the algorithm used to generate a query
$\bi{x}_k$, which induces a previously chosen local field $h_k$. The
solution, of course, depends on the weight vector $\bi{w}_k$ of the
hidden unit. This task is similar to solving a knapsack problem
\cite{Schroeder:1986:NTS}, which can be very difficult. But we need
only a fast approximate solution in order to use queries in the neural
key-exchange protocol.

As both weights $w_{k,j}$ and inputs $x_{k,j}$ are discrete, there are
only $2L+1$ possibilities for $w_{k,j} \cdot x_{k,j}$. Therefore we
can describe the solution by counting the number $c_{k,l}$ of products
with $w_{k,j} \cdot x_{k,j}=l$. Then the local field is given by:
\begin{equation}
  h_k = \frac{1}{\sqrt{N}} \sum_{l=1}^{L} l (c_{k,l} - c_{k,-l}) \, .
\end{equation}
We also note that the sum $n_{k,l} = c_{k,l} + c_{k,-l}$ is equal to
the number of weights with $|w_{k,j}|=|l|$. Hence the values of
$n_{k,l}$ depend only on the weight vector $\bi{w}_k$. This can be
used to write $h_k$ as a function of only $L$ variables, because the
generation of queries cannot change $\bi{w}_k$:
\begin{equation}
  h_k = \frac{1}{\sqrt{N}} \sum_{l=1}^{L} l (2 c_{k,l} - n_{k,l}) \, .
\end{equation}

In our simulations we use the following algorithm to generate the
queries. First the output $\sigma_k$ of the hidden unit is chosen
randomly. Therefore the set value of the local field is given by $h_k
= \sigma_k H$. Then we use either
\begin{equation}
  \label{eq:cl1}
  c_{k,l} = \left\lfloor \frac{n_{k,l} + 1}{2} + \frac{1}{2 l} \left(
      \sigma_k H \sqrt{N} - \sum_{j=l+1}^{L} j (2 c_{k,j} - n_{k,j})
    \right) \right\rfloor
\end{equation}
or
\begin{equation}
  \label{eq:cl2}
  c_{k,l} = \left\lceil \frac{n_{k,l} - 1}{2} + \frac{1}{2 l} \left(
      \sigma_k H \sqrt{N} - \sum_{j=l+1}^{L} j (2 c_{k,j} - n_{k,j})
    \right) \right\rceil
\end{equation}
to compute the values of $c_{k,L}$, $c_{k,L-1}$, \dots, $c_{k,1}$. In
each calculation one of the two equations is selected randomly with
equal probability, so that rounding errors do not influence the
average result. Additionally, we have to take into account that $0
\leq c_{k,l} \leq n_{k,l}$. Therefore we set $c_{k,l}$ to the nearest
boundary value, if \eref{eq:cl1} or \eref{eq:cl2} yield a result
outside this range.

Afterwards the input vector $\bi{x}_k$ is generated. Inputs associated
with zero weights are chosen randomly, because they do not influence
the local field. The other input bits $x_{k,j}$ are divided into $L$
groups according to the absolute value $l=|w_{k,j}|$ of their
corresponding weight. In each group, $c_{k,l}$ inputs are selected
randomly and set to $x_{k,j} = \mathrm{sgn}(w_{kj})$. The remaining
$n_{k,l} - c_{k,l}$ input bits are set to $x_{k,j} =
-\mathrm{sgn}(w_{k,j})$.

Simulations show that the absolute local field $|h_k|$ matches its set
value $H$ on average. And we observe only small deviations, which are
caused by the restriction of the input values to $+1$ or $-1$. So we
can generate queries, which approximately induce a predetermined
absolute local field $H$ by using this algorithm.

\section*{References}

\bibliographystyle{jstat}
\bibliography{paper}

\end{document}